# Underground operation at best sensitivity of the mobile LNE-SYRTE Cold Atom Gravimeter

T. Farah[1], C. Guerlin[1,‡], A. Landragin[1], Ph. Bouyer[2],
S. Gaffet[3], F. Pereira Dos Santos[1] and S. Merlet[1]

[1]*LNE-SYRTE, Observatoire de Paris, LNE, CNRS, UPMC, 61 avenue de l'Observatoire, 75014 Paris, France,*
[2]*LP2N, UMR5298, IOGS, CNRS, Université de Bordeaux 1, 351 cours de la Libération, 33405 Talence cedex, France,*
[3]*LSBB, UMS UNS, UAPV, CNRS, 84400 Rustrel, France )*

*Low noise underground environments offer conditions allowing to assess ultimate performance of high sensitivity sensors such as accelerometers, gyrometers, seismometers... Such facilities are for instance ideal for observing the tiny signals of interest for geophysical studies. Laboratoire Souterrain à Bas Bruit (LSBB) in which we have installed our cold atom gravimeter, provides such an environment. We report here the best short term sensitivity ever obtained without any ground vibration isolation system with such an instrument: $10^{-8}$ m.s$^{-2}$ in 100 s measurement time.*

## INTRODUCTION

Progress in high precision geophysical measurements is intimately linked to the development of sensitive instruments. Atom interferometry offers a new concept for inertial sensors [1], among which absolute gravimeters are the first ones having recorded geophysical signals [2]. Atom gravimeters now achieve state of the art performances as assessed by the three last main international gravimeter comparisons [3, 4] where one such instrument, presented in this paper, was participating for the first time.

Besides allowing a high sensitive absolute measurement, the architecture of atom inertial sensors has shown to be robust and versatile: they do not need for calibration and they can be operated in very different places and environmental conditions such as in an airplane [5] or within a 500 m depth underground research facility (this paper). Cold atoms gravity sensors also extend the range of time and length scales for variation of accessible signals: their repetition rate goes up to 330 Hz [6] and the possibility of using several atom clouds at several positions in a common laser field [7] allows for long baseline gradiometry. This principle is at the heart of a innovative large scale gravitational antenna, MIGA[†], being currently set up in the low noise underground research laboratory of Rustrel (LSBB URL). This facility will rely on the quiet environmental conditions offered by the site, in which several atom interferometers will be deployed within two few-hundred-meters-long galleries; it will offer access to multi dimensional gravity gradients and higher order moments, in a site of interest for studying the interactions between environment, society, internal Earth, atmosphere, nearby universe and interfaces and developing instrumentation with ultimate sensitivity.

At present, atom interferometers performances have been limited by environmental noises such as tilts, ground vibrations and parasitic rotations for example. Particularly as for any DC accelerometer the sensitivity is limited by parasitic vibrations, especially when performing measurements in anthropized or industrial areas. Different strategies have been developed to reject ground vibrations with passive [9] and active isolation systems [2, 10, 11] or in strap down configuration without any isolation [12, 13]. As an alternative these measurements can be performed in low vibration noise environments such as deserts, disused mines [14] and other underground galleries. Ultimately, space in a dedicated or natural satellite may also offer a good platform to test the limits of atom interferometers [15, 16, 17, 18].

In this context and as a first step to the MIGA project, our cold atom gravimeter (CAG) has been installed in the LSBB in order to evaluate the benefits this instrument can take from operating in such a low background noise environment. This paper presents the results of this measurement campaign. In particular, the low noise environment allowed the CAG to reach an unprecedented short term sensitivity when operating without any vibration isolation system.

---

‡ Present adress: LKB, CNRS, UPMC, UMR 8552, 24 rue Lhomond, 75005 Paris, France
† MIGA: Matter wave-laser Interferometry Gravitation Antenna [8] http://www.materwave-antenna.org

# THE TRANSPORTABLE ATOM GRAVIMETER

The absolute gravimeter developed at LNE-SYRTE, named CAG-01 in International Comparison of Absolute Gravimeters (ICAGs), operates since summer 2009. It participated to the three last main international or european comparisons which occurred since then [3, 4] respectively ICAG'09 at BIPM (France) and ECAG'11 and ICAG'13 at Walferdange (Luxembourg). Between these two first comparisons, the optical bench [19] and the two racks housing the control electronics have been combined in a single control ensemble allowing to move the gravimeter more easily. The instrument is now composed of two parts, the gravimeter core and the optical and electronic bench (Figure 1).

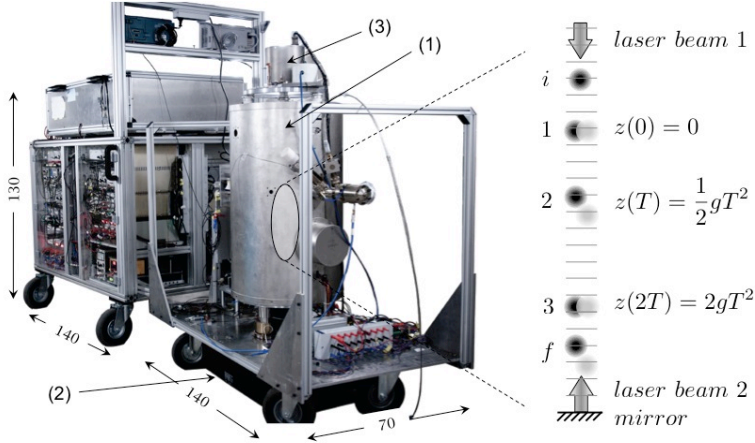

**Fig. 1.** Left, picture of the CAG. Forefront: the vacuum chamber is located inside the cylindrical magnetic shields (1), it lies on an isolation platform (in black) (2), a seismometer is attached to the chamber (top (3)). Second rate: the optical bench and electronics. Dimensions are in cm. Right, enlarged scheme of the free-fall section representing the successive positions of the atoms with respects to the referential frame defined by the two laser beams: i) initial trap position, 1) 2) 3) position at each pulse and f) final detection position.

As presented in [20], the CAG test mass is a $^{87}$Rb cold atomic sample at 2 μK which falls under gravity after the 3D-MOT lasers beams are adiabatically shut off. During the free fall, a sequence of three stimulated Raman transitions (π/2-π-π/2) separates along two paths and recombines the atomic wave function, using a pair of vertical lasers of frequencies $\omega_1$ and $\omega_2$ and wavelength $\vec{k}_1$ and $\vec{k}_2$. These lasers induce a two photon transition that couples the two hyperfine sublevels |F=1> and |F=2> of the $^5S_{1/2}$ ground state. The interferometer phase shift ΔΦ is derived from the measurement of the populations of the two output states thanks to the state labeling method [21]. The obtained transition probability is given by $P = \frac{1}{2}(1 + C \cos \Delta\Phi)$ where $C$ is the interferometer contrast. The cycling rate is 360 ms.

Each time the atoms experience a light pulse, the phase difference of the Raman lasers $\varphi$ is imprinted onto the atomic phase. This phase depends on the position of the atoms with respect to the surfaces of equal phase difference between the lasers, which provide a fine ruler for the measurement of the displacement of the atoms. Finally, the phase shift between the two paths I and II depends on $g$ the gravity acceleration, $\vec{k}_{eff}$ the wave vector difference between the lasers ($\vec{k}_{eff} = \vec{k}_1 - \vec{k}_2$), and scales with the square of the time interval between two consecutive pulses T [22]:
$\Delta\Phi = \Phi_{II} - \Phi_I = \varphi(0) - 2\varphi(T) + \varphi(2T) = \vec{k}_{eff}.\vec{g}T^2$.

The two laser beams are delivered to the atoms through a single collimator and are reflected by a mirror placed at the bottom of the UHV chamber. This architecture produces four beams onto the atoms [20], which allows driving the interferometer either with co- or counter-propagating transitions. Gravity measurements are performed using counter-propagating wave-vectors. Due to conservation of angular momentum and to the Doppler shift induced by the free fall of the atoms, only two of the four beams are resonant and drive the Raman transitions. Nevertheless, changing the polarisation configuration of the Raman beams and tuning the frequency difference between the Raman lasers on resonance with the hyperfine transition allows performing the interferometer with co-propagating wave-vectors. This reduces the sensitivity to $g$ and parasitic vibrations by 5 orders of magnitude while keeping the sensitivity of the interferometer phase $\Delta\Phi$ to all other sources of noise. Another advantage of the four beams Raman geometry lies in the possibility of choosing, by adjusting the sign of the Doppler shift between the two pairs of beams to realize the counter-propagating Raman transitions with a wave-vector pointing either upwards ($k_\uparrow$) or downwards ($k_\downarrow$).

The Raman lasers are phase-locked onto a low phase noise microwave reference source. Their frequency difference is swept according to $\omega_2 - \omega_1 = \omega_2(0) - \omega_1(0) + \alpha t$ in order to compensate for the Doppler shift induced by gravity. This adds a term $\alpha T^2$ to the interferometer phase, which cancels it for a perfect Doppler compensation. The central fringe of the interferometer thus corresponds to $\alpha_0 = \vec{k}_{eff}\cdot\vec{g}$ (Figure 2). The value of $g$ is therefore derived from the value of the frequency chirp $\alpha_0$ and the effective wave-vector.

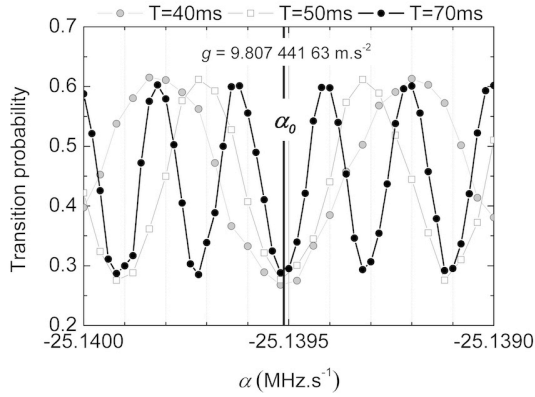

**Fig. 2.** Interferometer fringes obtained at LSBB by scanning the frequency chirp $\alpha$ for different $T$. The center fringe $\alpha_0$, whose position is independent of $T$, corresponds to the chirp rate that compensates gravity acceleration.

As maximal sensitivity to phase fluctuations is achieved when operating the interferometer at half fringe ($\Delta\Phi = \pm\pi/2$), the Raman phase is modulated by $\pm\pi/2$ so that the measurement is always performed at half fringe height, alternately on both sides of the central fringe. From two consecutive measurements of the transition probability $P_i$ and $P_{i+1}$, the phase error can be estimated. In practice, a correction $G\times(P_i-P_{i-1})$ is added at each cycle to $\alpha$, in order to stir the chirp rate onto the central fringe [12]. This realizes an integrator.

The signal to noise ratio is limited by any temporal fluctuations of the laser phase difference. Noise contribution arising from the lasers phase noise is reduced by phase-locking the laser phase difference. Among the many other sources of noise that affect the sensitivity, ground vibrations are dominant [9]. To reduce their influence, the gravimeter chamber lies on top of a passive isolation platform (Minus K 650BM-1 [23], vertical natural frequency of 0.5 Hz). Moreover, the interferometer phase $\Delta\Phi$ is corrected from the effect of the remaining vibration noise, which is measured independently with a low noise seismometer (Guralp T40) rigidly attached to the vacuum chamber (Figure 1). The complete vibration rejection scheme is detailed in [9, 12], an illustration of the correction vibration process is represented in figure 3. In particular we have demonstrated in [9] that the rejection efficiency is limited by the transfer function of the seismometer.

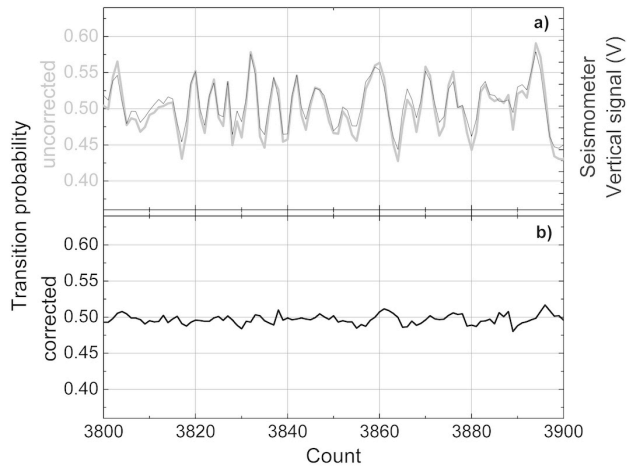

**Fig. 3.** Transition probability of the interferometer at LSBB without the isolation plateform. (a) bold light gray line represents the raw transition probability, dark gray line represents the seismometer vertical signal. The two signals are correlated. Using the seismometer signals, the transition probability is corrected and represented in bold black in (b).

**MEASUREMENTS CONDITIONS AT LSBB**

The *Laboratoire Souterrain à Bas Bruit* (LSBB URL) is located in the city of Rustrel in the South of France. It used to be one of the national ground based nuclear missile infrastructure. It is now a cross-disciplinary underground laboratory [24]. It takes advantages of its remote location and its robust design (it was built to remain operational even in the case of a nuclear blast). It allows to perform scientific measurements in a ultra-low noise environment. In particular, several seismometers from the 3D RUSF [25, 26] seismic antenna are installed at LSBB. The goal of the whole installation was to protect the firing command post "capsule", a Faraday cage, located in the deepest area of the 3.5 km tunnel, more than 500 m below the surface. Close to this capsule, the 10 m$^2$ R10 room has been arranged to place the CAG. To compensate for uneven ground, the drop chamber of the CAG has been installed on a dedicated concrete pillar 1.6 m × 1.6 m × 0.17 m (Figure 4).

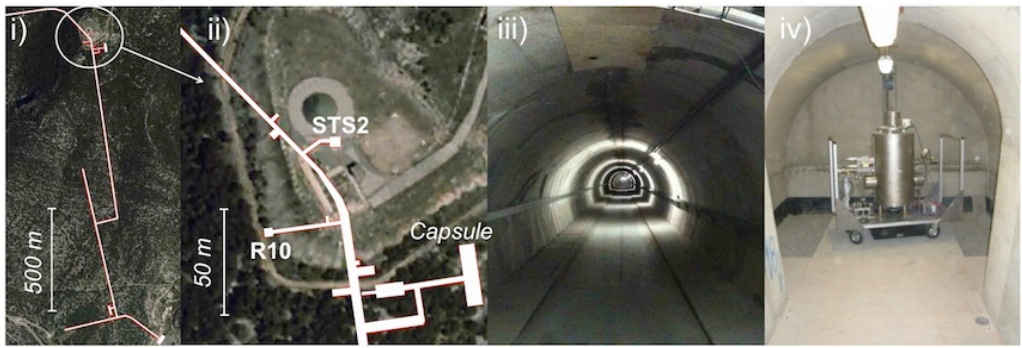

**Fig. 4.** Pictures of LSBB. From left to right, i) Aerial photography of the mountain where the LSBB is located, ii) Zoom of the R10 room area where the CAG was installed. 50 m away, a STS2 seismometer (RUSF.01) records ground vibrations continuously. iii) Picture of the LSBB tunnel and iv) CAG drop chamber installed in R10 room.

Figure 5 displays in bold the vertical vibration spectra measured with the CAG seismometer during the day (a et b traces), with and without the isolation platform. A broad resonance appears at 0.5 Hz when using the passive platform corresponding to its resonance frequency. Surprisingly, except at 5 Hz, 7 Hz and 10 Hz, we do not observe a significant reduction in the noise at higher frequencies that the platform is supposed to filter. We checked that the measurements are not limited neither by the noise of our FFT analyzer (HP 3561A) nor by the intrinsic noise of the seismometer (which starts dominating above only above 40 Hz). We attribute this noise to the acoustic noise and air flow generated by the CAG control electronics (computer fans, regulated power supplies ...)[‡]. Their influence is minimized by separating them from the drop chamber by the maximal distance allowed by the length of the optical fibers (of about 2 m) that connect the control electronics to the CAG chamber. Doing so we observed a significant reduction in the noise in the 1 Hz to 10 Hz band.

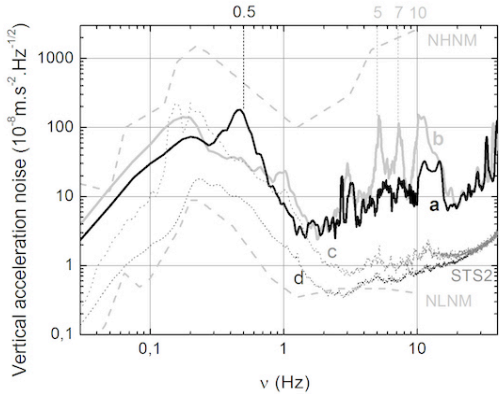

**Fig. 5.** Vertical acceleration noise spectra at LSBB. The *a)* (respectively *b)*) trace displays the spectrum obtained with (respectively without) the isolation platform, measured at daytime with the CAG seismometer. The thin spectrum *c)* has been obtained at the same moment with a STS2 seismometer (RUSF.01) located nearby, the *d)* spectrum was obtained on May 2012 and shows even better vibration conditions. Peterson models are also plotted.

A spectrum obtained the same day with the RUSF.01 seismometer (STS2) located 50 m away (*c*) trace in figure 5) is similar at low frequency but insensitive to the acoustic noise generated in the CAG room. This vibration noise level is not the lowest one can encounter in this place. The *d)* trace on Figure 5 displays a spectrum recorded on May 2012 with the same STS2 which is significantly lower, close to the New Low Noise Model of Peterson [28].

**ATOMIC GRAVIMETER MEASUREMENTS**

For the measurements presented here we realized two interleaved integrations corresponding to opposite wave-vectors. In practice the direction of the $\vec{k}_{eff}$ wave-vector was reversed at each cycle. Combining the results of these two integrations we obtained a gravity measurements corrected from the phase shift associated to the one photon light shift as well as its fluctuations over timescales longer than the cycle time [2, 20]. The sensitivity to the power fluctuations of the Raman lasers is thus drastically reduced. With respect to the measurement protocol we routinely use to obtain accurate *g* measurements [20], we do not correct here for the effect related to the two photons light shift [29], which requires additional measurements that compromise the short term sensitivity.

---

[‡] To shield the CAG drop chamber from acoustic noise in our laboratory [27], we usually surround it with a wood box covered with dense isolation acoustic foam. This box being too large to fit on the concrete pillar was not used.

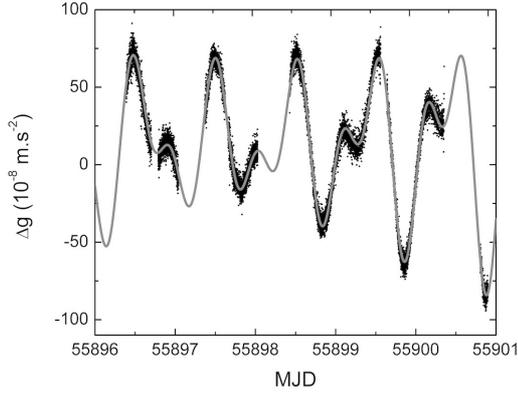

**Fig. 6.** CAG gravity measurements at LSBB uncorrected from tides. One black dot corresponds to two drops (720 ms). Tidal prediction variation is represented in gray.

Figure 6 shows examples of measurements obtained with the isolation plateform, performed during nights and week-end with the two interleaved integrations with opposite wave-vectors leading to a cycling time of 2×360 ms. On figure 6, each data point corresponds to 720 ms. As the passive isolation platform resonant frequency is 0.5 Hz, it thus efficiently filters vibration noise above about 1 Hz, at the expense of a slight increase of the noise in the 0.1 Hz to 1 Hz band. As the vibration noise above 1 Hz is very low at the LSBB (Figure 4), measurements without any isolation platform have also been performed. Results are presented on Figure 7 which displays Allan standard deviations of gravity fluctuations. This gives the statistical uncertainty as a function of the averaging time. The four traces correspond to measurement with (a) or without isolation platform (b), with (in black) or without post-correction scheme (in gray).

First, the black trace in a) presents the results in our usual operating mode: with isolation platform and post-correction scheme. For short averaging times the Allan standard deviation is filtered due to the time constant of the integrator of the order of a few cycles but still displays bumps which tend to average which we attribute to the swell. After 20 s the Allan standard deviation decreases with integration time as $\tau^{-1/2}$ which is characteristic of a white measurement noise. It reaches $10^{-8}$ m.s$^{-2}$ after 100 s only, decreases down to a plateau at $7\times10^{-9}$ m.s$^{-2}$. This long-term stability is comparable to the one we demonstrated in [20]. The associated gray trace displays the vibration noise derived from the 3D-seismometer from which the gravity data are corrected. The difference between the two curves illustrates the efficiency of the vibration rejection by a factor ranging from 5 to 3 depending on the time scale.

Second, curves in b) display results obtained without isolation few minutes after. Remarkably the sensitivity on the gravity measurement (in black) is similar while the vibration noise is significantly higher (in gray). In particular, we find a much larger rejection efficiency of about 8, almost independent of the averaging time. Reaching such a low sensitivity without vibration noise isolation constitutes a remarkable result and demonstrates the quality of both the instrument and the LSBB site. This is a factor of two better than the sensitivity we obtain routinely in the LNE laboratory with the isolating platform and the surrounding anti-acoustic box. It is comparable to our best sensitivity of $14\times10^{-8}$ m.s$^{-2}$.Hz$^{-1/2}$ (obtained in the LNE-SYRTE laboratory in Paris by night with the air conditioning of the laboratory turned off [8]). It is a factor 2 above the best short term sensitivity ever reported of $4.2\times10^{-9}$ m.s$^{-2}$.Hz$^{-1/2}$ at Wuhan [30] (where a very efficient active isolation system based on a superspring had been implemented).

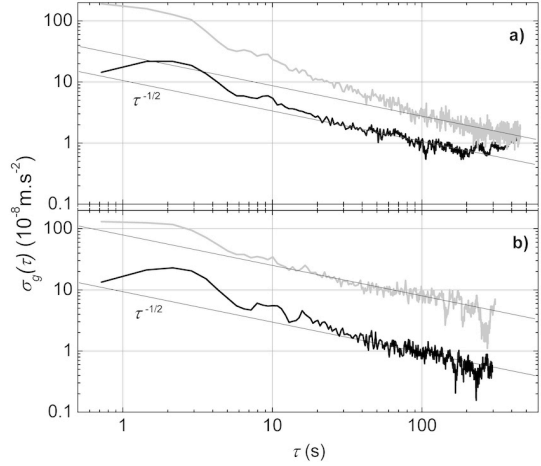

**Fig. 7.** Allan standard deviations obtained at LSBB with isolation platform (a) and without (b). Black curves: gravity sensitivities when post correcting from ground vibrations. Gray curves represent the sensitivities of the associated seismometer signals. $\tau^{-1/2}$ slopes correspond to the averaging of a white noise.

The main difference between the two measurements lies in the long term residual vibration noise level (gray traces on Figure 6). In particular, when using the isolation platform, the vibration noise averages faster than a white noise up to 100 s and reaches a lower level than when the platform is not used. In both cases, the noise averages to reach white noise behavior for long measurement times (scaling as $\tau^{-1/2}$). Aliasing effects can explain this: the sensitivity is degraded by vibration noise components at frequencies multiple of the repetition rate, 3 Hz in our case [31]. The comparably lower level of white noise we obtain when using the platform is the evidence that vibration noise at high frequency is indeed filtered by the passive platform. As for the short term part, it is dominated by the vibration noise in the 0.1 Hz to 1 Hz band, which is reasonably well sampled and thus averaged faster than white noise, thanks to our high repetition rate.

It is somewhat surprising that despite the fact that using or not the platform modifies the "color" of the vibration spectrum, we find equal level of sensitivities. This tends to indicate that we reached here the intrinsic limit of our instrument.

Using the spectrum of the seismometer self noise provided by the manufacturer, we calculated a contribution to the gravity measurement smaller than the sensitivity we report here, of $4\times10^{-8}$ m.s$^{-2}$ after 1 s measurement time. Independent measurements of this self noise could be useful to assess this limit when using the seismometer to correct gravity data. Measurements with other sismometers would also be interesting

The "self-noise" of the atom gravimeter itself has been evaluated independently (excepted vibration). The contribution to the gravity sensitivity to gravity is estimated to be at the level of about $3\times10^{-8}$ m.s$^{-2}$ at 1 s, significantly smaller than the level we obtain here˅.

We cannot exclude that reaching the same level of performance with or without the platform is simply a coincidence. Vibration noise spectra with and without platform do have some differences, and the rejection efficiency, which is limited by the response function of the seismometer, is frequency dependent [12]. Another possibility would be that this behavior is related to the averaging of the seismometer signal at frequencies below its corner frequency, which could explain similar Allan standard deviations for long averaging time.

---

˅ This estimation results from measurement in co-propagating configuration and the contribution of the master laser frequency noise which was shown in [9] to be a limit at the level of $4\times10^{-8}$ m.s$^{-2}$ at 1 s.

## SUMMARY AND OUTLOOK

We have performed at LSBB measurements with a mobile atom gravimeter reaching a record sensitivity of $10^{-8}$ m.s$^{-2}$ in only 100 s measurement time. This was achieved without any vibration isolation system, but with the correction of the vibration noise based on simultaneous seismic ground motion measurements. Such a sensitivity would allow in principle to perform the measurements during the ICAGs in a single day only, as the usual protocol requires three independent measurements at three different stations. As a comparison with the measurement protocole usually used with free falling corner cube, this sensitivity corresponds to a set scatter of $10^{-8}$ m.s$^{-2}$ for each set of only 100 s.

Remarkably, the low frequency vibration noise averages faster than white noise, thanks to the high repetition rate of our instrument. This is a clear advantage of cold atom gravimeters with respect to free falling corner cube gravimeters which operate at a lower repetition rate, resulting in undersampling of the microseismic noise.

For the measurements presented here, the electronics control system of our gravimeter was found to be a significant source of high frequency noise. This noise is nevertheless efficiently corrected for, and its influence is reduced by the use of the passive isolation platform. Its effect could be reduced further by separating the control system further from the vacuum chamber, using longer optical fibers, and/or with a more careful design reducing the generation of acoustic noise. Also, a surrounding acoustic enclosure around the drop chamber, such as the wooden box we usually use, would probably allow to take full advantage of the LSBB quietness.

In conclusion, the excellent noise conditions we met at LSBB make this laboratory suitable for high sensitivity gravity measurements. The large size of this infrastructure makes it perfectly suitable for the organization of large gravimeter intercomparison campaigns, as well as for the implementation of a large scale gravitational antenna. This last infrastructure, MIGA, will be deployed in the next five years at LSBB and will use a set of atom interferometers to measure gravity gradients over a long baseline. The determination of such gradients will be based on differential measurements, providing an additional suppression of the common mode vibration noise, and will allow to reach unprecedented sensitivities in strain measurements.

## ACKNOWLEDGMENTS

The measurement campaign has been supported by Action Spécifique GRAM (Gravitation Références Astronomie et Métrologie). The authors would like to thank Christophe Sudre, Michel Auguste, Alain Cavaillou, Julien Poupeney and Daniel Boyer of LSBB for their invaluable help throughout the measurement campaign.